\documentclass[amssymb,twocolumn,
pra,aps,floatfix]{revtex4}
\usepackage[paperwidth=210mm,paperheight=297mm,centering,hmargin=1.7cm,vmargin=1.9cm]{geometry}
\usepackage{graphicx}
\usepackage{dcolumn}
\usepackage{bm}
\begin{document}

\title{Pulse-shape control of two-color interference in high-order-harmonic
generation}

\author{K. R. Hamilton}%

\author{H. W. van der Hart}%
\author{A. C. Brown}%
\affiliation{Centre for Theoretical Atomic, Molecular and Optical Physics, Queen's University Belfast, Belfast  BT7 1NN, United Kingdom
}%

\begin{abstract} 

We report on calculations of harmonic generation by
neon in a mixed (800-nm + time-delayed 400-nm) laser pulse scheme. 
In contrast with previous studies we employ a short (few-cycle) 400-nm pulse,
finding that this affords control of the interference between electron
trajectories contributing to the cutoff harmonics.
The inclusion of the 400-nm pulse enhances the
yield and cutoff energy, both of which 
exhibit a strong dependence on the time delay between the two pulses. 
Using a combination
of time-dependent $R$-matrix theory and a classical trajectory model, we
assess the mechanisms leading to these effects.

\end{abstract}

\maketitle

\section{\label{sec:intro}Introduction}

The process of high-order-harmonic generation (HHG) has been the driving force behind
countless new developments in ultrafast laser technologies over the past decade. HHG has
been used to create both 
short-duration \cite{67as_pulse} and high-energy \cite{efficient_kev_hhg} laser pulses and can also be
used directly in atomic \cite{atom_hhg} and molecular
spectroscopy \cite{tomographic_molecular} to elucidate the attosecond-scale 
dynamics of electrons. 

HHG is commonly described by the classical three-step model in which an
electron (i) tunnels through the laser-suppressed Coulomb barrier, (ii) is
accelerated by the field, and (iii) recombines with its parent atom emitting a
high-energy photon, all within a single cycle of the driving laser field
\cite{corkum1993}. Analysis of the electron motion reveals
two classes of trajectory. The so-called
long and short trajectories represent two distinct pathways leading to the same
recollision energy. 
Importantly, the cutoff (highest) energy trajectories arise when the long and
short trajectories coalesce. It has been shown that interference between these pathways 
leads to an Airy pattern in the harmonic plateau \cite{airy_harmonics}.

The study of spectral caustics in HHG is an extension of this interference
effect.
 Caustics arise in the analysis of ray optics: where multiple rays
coalesce a sharp peak, or caustic, appears in the emitted radiation. This enhancement is predicted by
catastrophe theory, and is due to a singularity in the spectral density. In HHG,
caustics result from the coalescence of more than two electron trajectories in
the same spectral region, which can be engineered by adding a second color to
the driving laser pulse. The second color breaks the symmetry of the process,
splitting the electron trajectories into two further classes. The so-called
upper and lower branch trajectories (named as they yield higher or lower cutoff
energies than the equivalent one-color trajectories) can then interfere and, when they coalesce
at the cutoff energy, yield a dramatic enhancement in the harmonic spectrum
\cite{spectral_caustics_second_harmonic, spectral_caustics_giant_resonance}.

Such interference between trajectories represents an attractive means of probing the
quantum nature of the HHG process, which has been understood primarily as a
strong-field (classical) process. However, this is experimentally challenging,
because the measured harmonic spectrum arises from the coherent response of many
atoms and, depending on the experimental conditions, only either the short or long
trajectories can be appropriately phase matched. Additionally, in a two-color
field, the upper branch
trajectories have a reduced ionization and recollision probability relative to
the lower branch. Recently, however, it was shown
that a shape resonance in the harmonic spectrum can compensate for these
factors, and reveal the spectral interference of the relevant trajectories
\cite{spectral_caustics_giant_resonance}.

The manipulation of electron trajectories with two color fields has been
realized in various schemes.
Schafer and coworkers proposed the use of a combined infra-red (IR) / extreme
ultraviolet (XUV) scheme to control the electron trajectories
\cite{path_control_efficiency}. More recently the XUV-initiated HHG scheme has
been applied to monitor core-hole dynamics in small molecules
\cite{correlation_core_hole_dynamics}, and to elucidate the contribution of both
inner and outer valence electrons to the HHG spectrum \cite{brown_xuvhhg}.

A more established technique involves the use of two colors in the visible-IR
range, and in particular using a driving pulse and its second harmonic ($\omega
+ 2\omega$) has been well studied.  The inclusion of the second harmonic has
been shown to enhance the high-harmonic yield \cite{2harm_enhancement} and extend
the cutoff energy of the harmonic spectrum \cite{2colour_extension}. Thus,
two-color fields have been used extensively for the generation of supercontinua
in the XUV range \cite{xuv_supercontinuum,supercontinuum}, the spectral
shaping of attosecond pulse trains \cite{atto_2colour_shape}, and quantum path
selection in HHG \cite{2colourexpt}. 

In all previous studies, the
second harmonic is included as a long-duration dressing
field which imparts some phase-dependent effect to the harmonic spectrum.
In the
present paper we will instead consider the interference between electron
trajectories driven by two few-cycle pulses. 


We obtain the harmonic spectra using the \emph{ab initio} time-dependent
\emph{R}-matrix method known as RMT \cite{rmt_paper}. RMT has been used to investigate various
strong-field phenomena, including multielectron correlation in doubly and
core-excited states in Ne \cite{RMT_ATAS}, electron recollision in F$^-$
\cite{ola_rescattering}, IR-assisted photoionization of Ne$^+$
\cite{RMT_rydberg} and HHG from noble gas atoms in the NIR regime
\cite{brown_xuvhhg,ola_nearIR}. A predecessor to the RMT method was also used
extensively to study harmonic generation in a variety of targets in the
UV--visible range \cite{brown_prl,ola_multichannel_ne+,brown_ar+,brown_m1}.  The
RMT approach has two defining capabilities. First, the code is optimized to run
on massively parallel (\textgreater 1000 cores) machines, thus making the
extension to challenging physical systems tractable. Second, the RMT approach
can be applied to general multielectron systems\textemdash{} including open-shell
atoms and ions\textemdash{} with a full description of electron correlation effects.

Here we use the {\it ab initio} RMT method to apply a quantitative
analysis to spectral caustics in two-color HHG schemes.
We first give an overview of the RMT method and the
calculation parameters employed, then present results of calculations
of the harmonic spectrum for a neon atom in a combined 800/400-nm pulse scheme.

\section{\label{sec:tdrm}Time-Dependent $R$-Matrix Theory}

The $R$ matrix with time-dependence method (RMT) employs the well-known $R$-matrix
paradigm of 
dividing configuration space into two separate regions, in this case over the
radial coordinate of an ejected electron. In an inner
region (close to the nucleus) all electron-electron
interactions are taken into account while in the outer region an ejected
electron is sufficiently far from the residual ion that the effect of electron-exchange can
be neglected. The most appropriate numerical method for solving the
time-dependent Schr\"{o}dinger equation (TDSE) is employed in each region and, at variance
with other $R$-matrix-based approaches, the wavefunction itself is matched
explicitly at the
boundary rather than via an $R$ matrix \cite{rmt_paper}.

In the inner region the time-dependent $N$-electron wave function is represented
over an $R$-matrix basis with time-dependent coefficients. The basis is
constructed from the $N-1$ electron-states of the residual ion coupled to a complete
set of single-electron functions representing the ejected electron. Additional
$N$-electron correlation functions can be added to the basis set to improve the
accuracy of the wave function. In the outer region, the wave function is described
in terms of residual-ion states coupled with the radial wave function of the
ejected electron and is expressed explicitly on a finite-difference grid. The
two region wav efunctions are then matched directly at the boundary in two
directions.  The outer region finite-difference grid is extended into the
inner region and the inner region wave function is evaluated on this grid. This
provides the boundary condition for the solution of the TDSE in the
outer region. A derivative of the outer region wave function at
the boundary is also made available to the inner region, enabling the inner
region wave function to be updated. The wave function is propagated in the length gauge, as it has been found to give better results with the atomic structure description employed in time-dependent R-matrix calculations \cite{length_gauge}.

The harmonic spectra are obtained by evaluating the time-dependent expectation
value of the dipole velocity, then Fourier transforming and squaring it, thus:
\begin{equation}
    \mathbf{\dot{d}}\left(t\right)=\frac{d}{dt}\langle \Psi \left(t\right)
  |-e \mathbf{z}|\Psi\left(t\right)\rangle,
  \nonumber
  \end{equation}
where {\bf z} is the total position operator along the laser polarization axis
and $\Psi$ is the wave function.
It is also possible to calculate the spectra using the expectation value of the
dipole length \cite{brown_helium}, and within our simulations this is used as a
check for accuracy.  Indeed the spectra produced from each method show excellent
agreement with each other until well past the cutoff energy. The spectra shown
in the present paper
are those calculated from the dipole velocity.

To aid our analysis of the harmonic spectra we perform classical trajectory simulations based on the three-step model \cite{corkum1993}. We assume an electron is tunnel ionized into the continuum with zero initial velocity, and, for each possible ionization time, we determine the electron’s velocity and position by integrating over the acceleration in the two-color field. We note that this model does not account for any effect of the atomic potential. Trajectories which return again to the origin represent recolliding electrons which give rise to harmonic generation. The energy of this emmitted harmonic light can then be calculated from the electron’s recollision energy and the ionization potential. 

\section{\label{sec:calcparameters}Calculation Parameters}

The neon atom is described within an $R$-matrix inner region with a radius
of 20a.u. and an outer region of 2500 a.u. An absorbing boundary, beginning at 1500 a.u., is included to
prevent reflections of the wave function. The finite-difference grid spacing in the outer region is 0.08 a.u. and the time step for
the wave-function propagation is 0.01 a.u.  The description of neon includes all
available $2s^22p^5$$\epsilon \ell$ and $2s2p^6$$\epsilon \ell$ channels up to
a maximum total angular momentum of $L_{\mathrm{max}}=79$.
The inner region continuum functions are constructed using a set of 50 $B$ splines of order
9 for each available angular momentum of the outgoing electron.

We employ a mixed laser pulse scheme comprising a fundamental (800-nm) pulse and
its second harmonic (400-nm). In every mixed laser pulse scheme employed the pulses are linearly polarized and parallel.
The intensities of the 800 and 400~nm pulses are fixed at  $4\times10^{14}
\mbox{ W/cm}^2$ (corresponding to a
ponderomotive energy of 24 eV)  and $4\times10^{13}  \mbox{W/cm}^2$, respectively.
Both pulses employ a $\sin^2$ ramp on/off and are, unless otherwise stated, six
cycles in duration (three cycles ramp on, three cycles ramp off).
The time delay is measured between the central peaks
of the two pulses, and a negative time delay corresponds to the 
400-nm pulse arriving first. We increment the time delay in steps of 0.1~fs.

\section{\label{sec:results}Results}
\begin{figure*}
 \centering
  \includegraphics[width=1.0\textwidth]{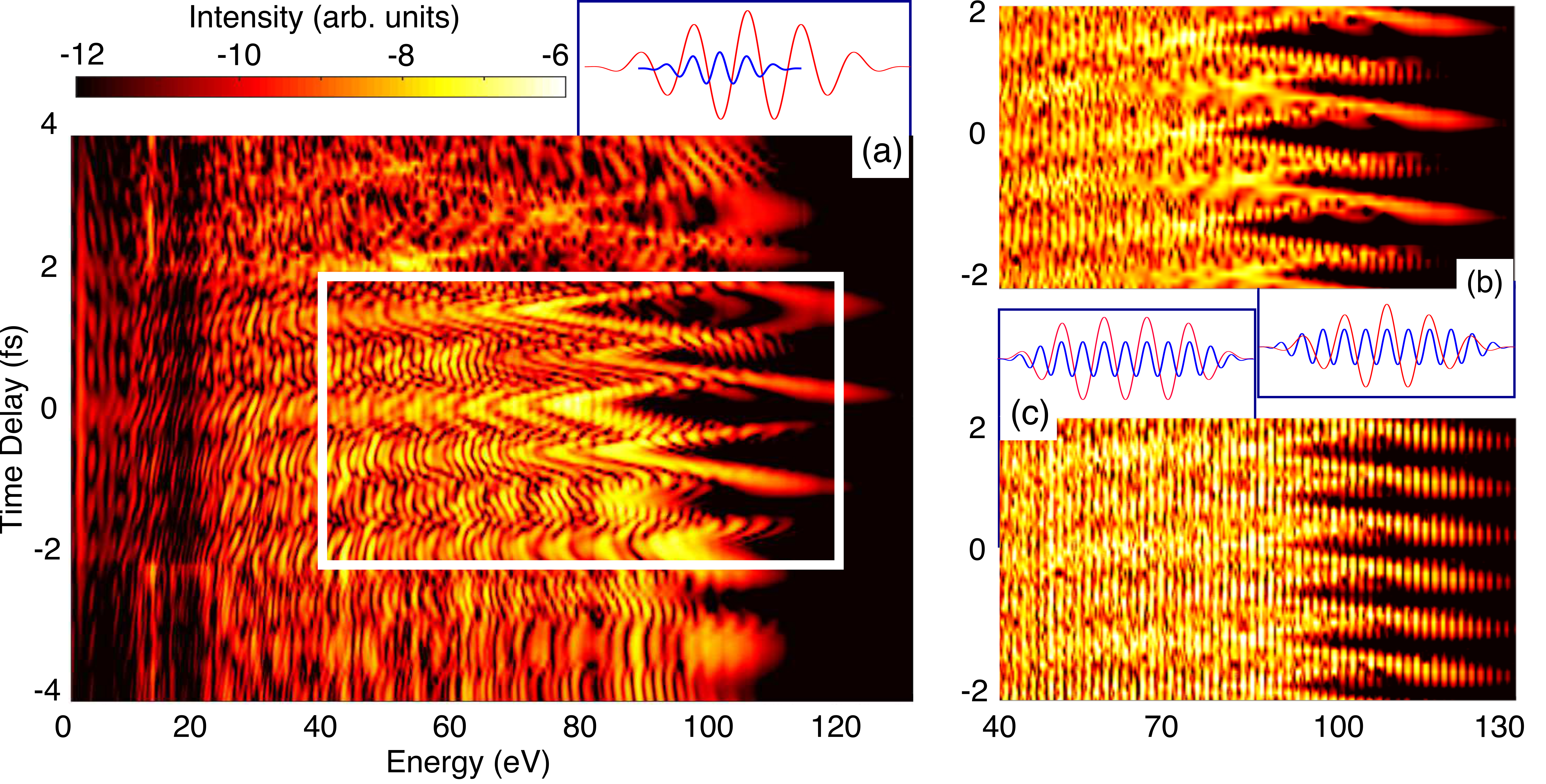}
  \caption{Harmonic spectra produced by neon irradiated by (a) a six-cycle, 800-nm
    pulse and time-delayed six-cycle,
    400-nm pulse; (b) a six-cycle, 800-nm pulse and a
    time-delayed 12-cycle, 400-nm pulse; and (c) a six-cycle, 800-nm pulse with two cycles at peak intensity
    and a time-delayed 12-cycle, 400-nm pulse. Insets show pictorially the two laser fields
    used. \label{short}}
\end{figure*}

Figure \ref{short}(a) shows the harmonic spectra obtained from neon irradiated by a six-cycle
800-nm ($\omega$) pulse and a time-delayed six-cycle 400-nm ($2\omega$) pulse. The inclusion of the
$2\omega$ pulse breaks the symmetry of the three step process, and stimulates
the generation of even harmonics \cite{even_harmonics}.
The two-color field also elicits an enhancement of up to four orders of magnitude in
the harmonic emission compared
to the primary field alone (not shown), particularly in the cutoff region
\cite{2harm_enhancement}. This is one to two orders of magnitude higher than
the enhancement achieved by simply increasing the intensity of the primary field
by 10$\%$ to match the combined peak intensity of the two-color field. 

This
enhancement is manifest most clearly in the appearance of the
distinctive swallowtail caustics in the
cutoff harmonics.
These sharp peaks in harmonic emission represent a singularity in the
harmonic emission, caused by the coalescence of multiple
electron trajectories in the two-color field
\cite{spectral_caustics_second_harmonic}. Changing the time delay between the $\omega$ and
$2\omega$ pulses changes the phase relationship between the interfering
trajectories, shifting the coalesence to higher energy and leading to the two
"arms" of the swallowtail. The series of swallowtails which peak at the
cutoff energy and decrease in
intensity with decreasing harmonic photon energy describe an Airy pattern
\cite{airy_harmonics}.

The swallowtail caustics are most clearly visible when two short pulses are used [Fig.
\ref{short}(a)], but the
overall yield is increased by the use of a longer $2\omega$ pulse. 
Figures \ref{short}(b) and \ref{short}(c) show the spectra produced using
a 12-cycle 400-nm pulse. In both cases the 400-nm pulse has
a three-cycle $\sin^2$ ramp on and off, with six cycles at peak intensity (3-6-3); in
Figure \ref{short}b the 800-nm pulse has a 3-3 form while in Fig. \ref{short}c
it has a 2-2-2 form. Using a long
$2\omega$ pulse is more similar in spirit to previous studies in this field,
where the second harmonic can be treated as a dressing field which imparts a
phase dependent enhancement or suppression on the harmonic yield
\cite{2harm_enhancement, 2colour_extension, xuv_supercontinuum, supercontinuum,
atto_2colour_shape, 2colourexpt}.  

\begin{figure}
  \centering
  \includegraphics[width=0.48\textwidth]{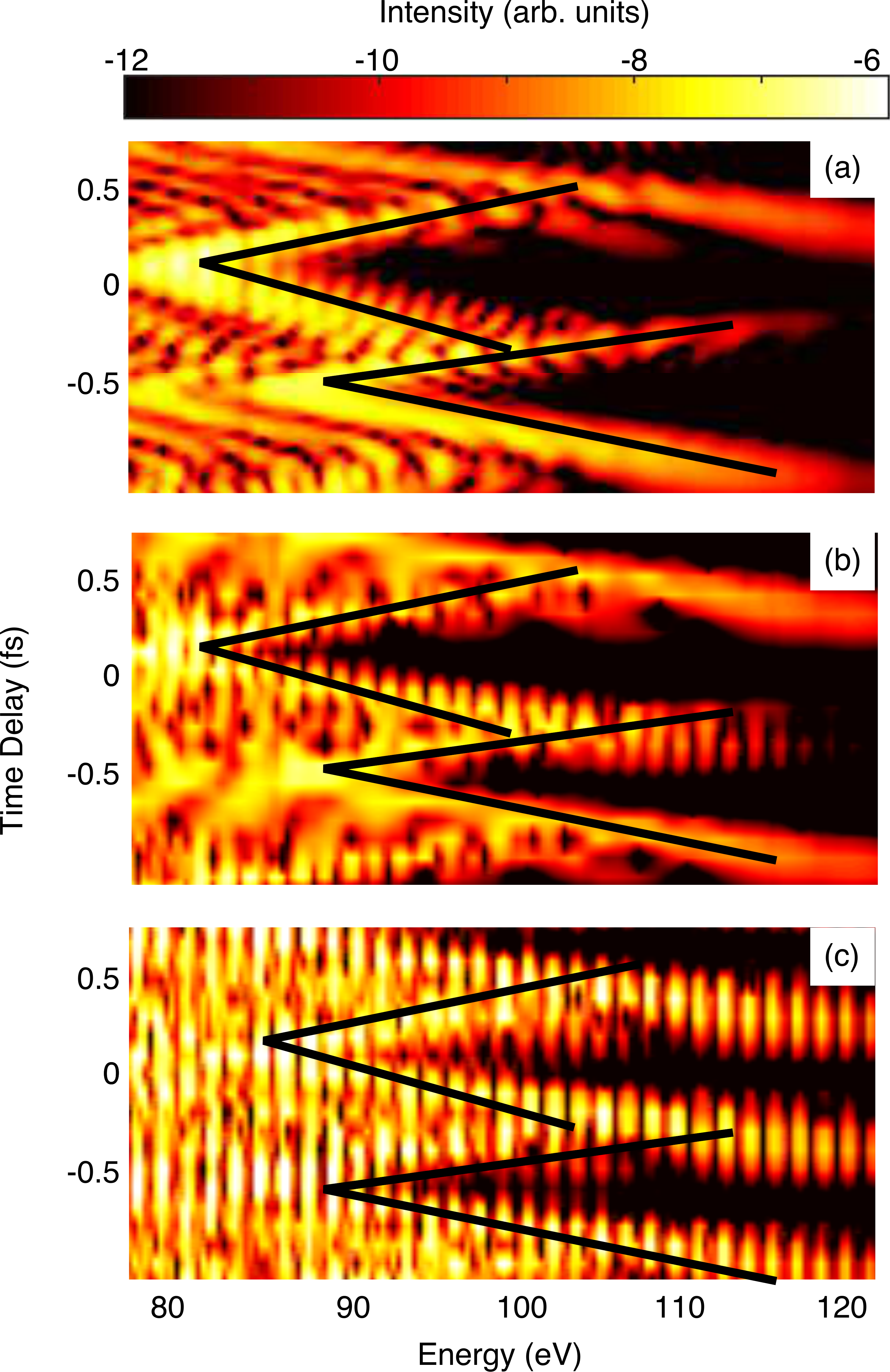}
   \caption{A zoomed view of the spectral caustics for each of the laser pulse
     scenarios described in the caption of Fig. \ref{short}. The swallowtail
     caustics are marked with black lines in each. \label{caustic_zoom}  
   }
\end{figure}

\begin{figure}[t]
  \centering
  \includegraphics[width = 0.45\textwidth]{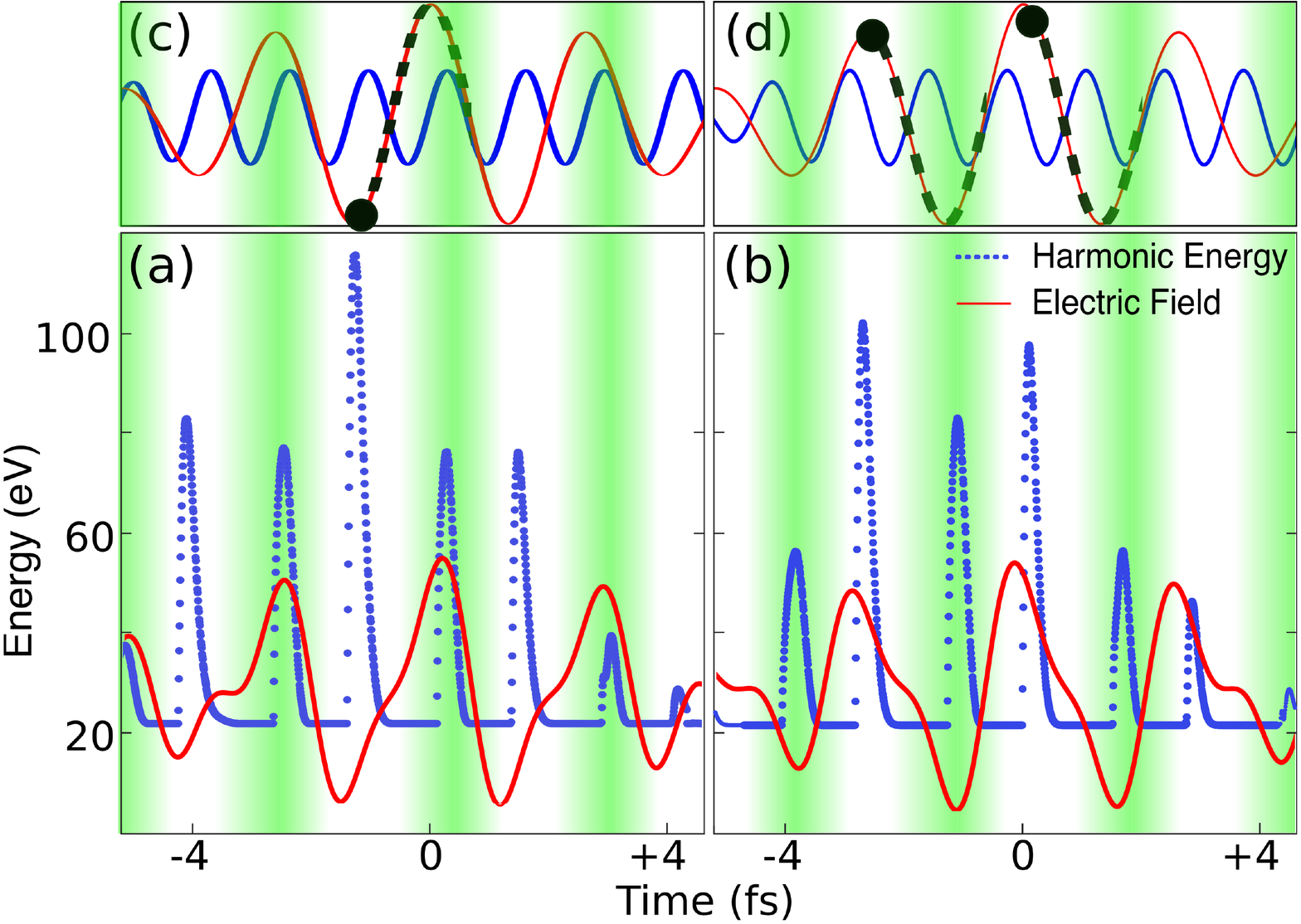}
  \caption{ The harmonic photon energy (blue dots) produced by electrons "born" at a given
    time during the laser pulse (red line) as calculated by a classical trajectory
    model, for a time delay of (a) 0.3~fs ($2\omega$ peak arriving just after the
    $\omega$ peak) and (b) 1.0~fs ($2\omega$ peak arriving just before the $\omega$ peak). Panel (a)
    generates one trajectory with the highest (cut-off) energy, while panel (b)
    generates two which interfere. The individual $2\omega$ and $\omega$ fields
    are shown in panels (c) and (d) where the position of the optimal electron emission
    times are marked by black dots. The green shaded regions correspond to
    the overlap of two peaks or two troughs, while the unshaded (white)
    regions correspond to the overlap of a peak and trough. Thus trajectories
    launched in the green regions do so with higher probability, as the
    tunnel-ionization probability is increased. \label{trajectories}}
\end{figure}

Using a long $2\omega$ pulse yields an increase in harmonic emission relative to the
short-$2\omega$-pulse spectra in Fig. \ref{short}(a). The $2\omega$ field
is present for the duration of the 800-nm pulse, and thus affects electron trajectories
originating earlier or later in the fundamental pulse. Thus
the emission is also less concentrated in the cutoff region such that the
entire plateau is enhanced. Furthermore, the interference patterns
in the time-delay spectra are strongly modified by the shape of the
pulses.  Changing from a 3-3 to a 2-2-2 pulse shape is similar to broadening the
carrier envelope of a Gaussian 800-nm field: with this profile there is no longer
a single central peak. Thus each harmonic energy is sourced by 
trajectories on multiple cycles, and the interference between them appears as the comblike fringes in
Fig. \ref{short}(c). The arms of the swallowtail caustic are still visible in the
cutoff region, but
the singular enhancement at the point of the caustics is somewhat bleached by the
overall increase in the harmonic yield. Changing the pulse profile also has the
effect of shifting the caustic to higher energy, as shown in Fig.
\ref{caustic_zoom}, because the trajectories leading to the caustic are now driven by
higher intensity cycles of the 800-nm pulse. For the same reason, the
swallowtail caustic which appears for a delay of $\approx -0.5$fs in Figs.
\ref{caustic_zoom}(a) and \ref{caustic_zoom}(b) is at higher energy than the main
caustic at $\approx +0.17$fs: the particular half cycle of the 800-nm pulse
which drives the
cutoff trajectories is stronger, as depicted in Fig. \ref{trajectories}(c)
compared with Fig. \ref{trajectories}(d).

Using a profile with a single
central peak yields two different interference structures in the cutoff
harmonics in Fig \ref{short}(b)\textemdash{} a long smeared out finger extending to high energy and a lower-energy comb-like
interference pattern.

We suggest the following interpretation of these interference structures in the
strong-field context.
The highest-energy trajectories are launched when a (peak) trough of the $2\omega$
field arrives just after a trough (peak) of the $\omega$ pulse i.e., when the
vector potentials of the two fields are oppositely oriented (unshaded regions in
Fig. \ref{trajectories}).
In a short pulse, the highest-energy trajectories are
launched at the trough one half cycle prior to the central peak (Fig. \ref{trajectories}(c):
black dot).
Therefore, for delays where a peak of the 400-nm pulse arrives just after the
penultimate trough of the 800-nm pulse, there is one trajectory with the highest
return energy, which manifests as broad harmonic peaks at the cutoff 
\cite{ola_nearIR}, and appears as the long smeared out fingerlike feature in
Fig. \ref{short}(b). 

By contrast, if troughs of the 400-nm pulse occur just after the penultimate and main peaks of the 800-nm
pulse, trajectories of approximately equal return energy are launched from each
(Fig. \ref{trajectories}(d): black dots).
Interference between harmonic light generated by the two
trajectories yields a comb-like interference pattern. This picture is supported
by classical trajectory calculations, which reveal that the highest-energy harmonics are sourced from one or two cycles of the fundamental pulse
depending on the time delay.

\begin{figure}[t]
  \centering
  \includegraphics[width=0.48\textwidth]{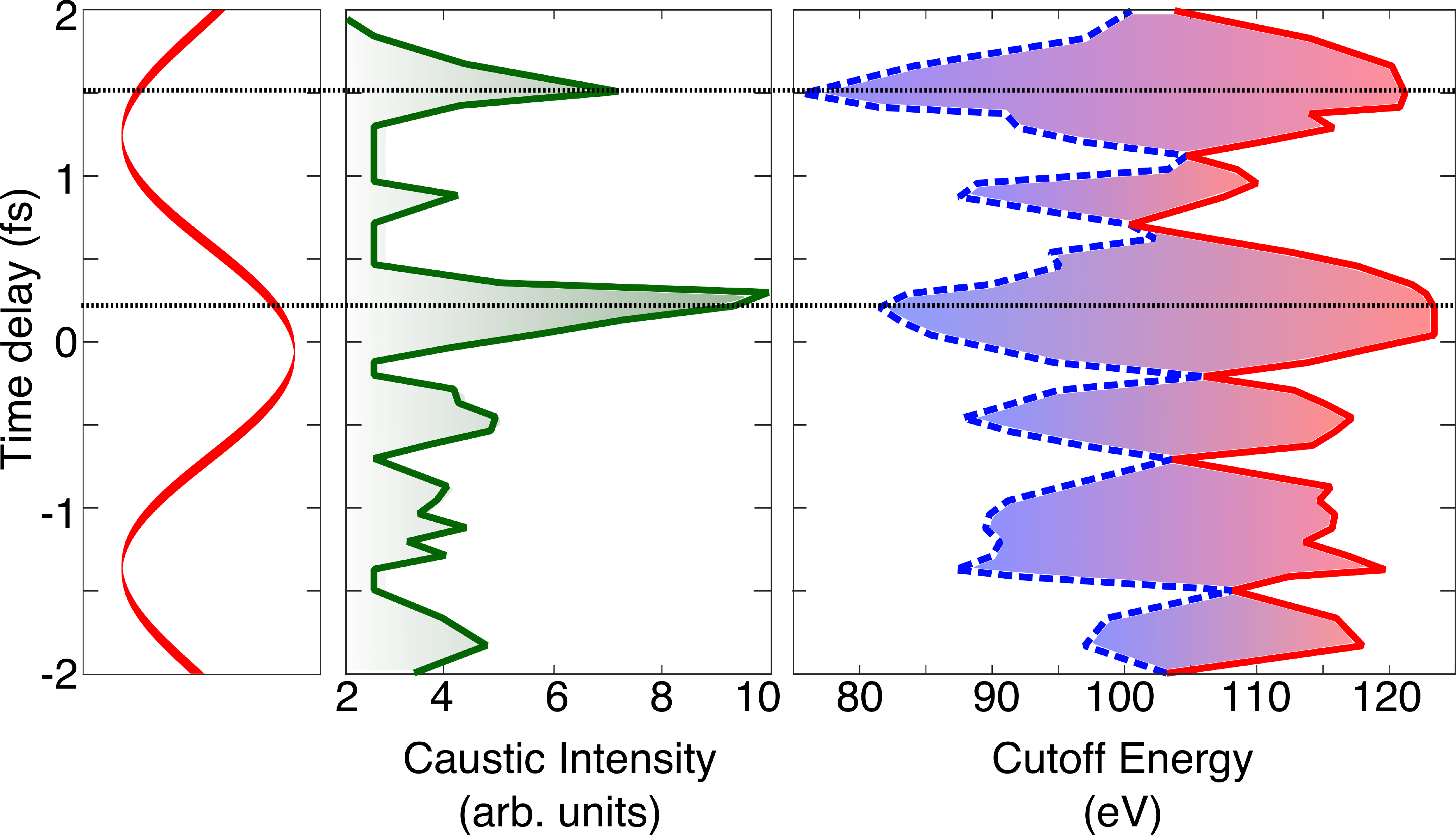}
  \caption{Intensity of the spectral caustic (central panel) and cut-off energy (right
    panel) from neon irradiated by short (six-cycle) 800- and 400-nm pulses with
    varying time delay. The cutoff of both the first (lower energy: blue dashed
    line) and second (higher energy: red solid line) plateaus are shown. The 800-nm pulse profile is shown for reference in the
    left panel. The time delays which give most intense caustic enhancement and the lowest or highest
    first or second cutoff are marked with the horizontal lines.
    \label{cutoffandyield}}
\end{figure}

Figure \ref{cutoffandyield} shows the cutoff energy and
caustic intensity as a function of time delay for the short $2\omega$ pulse
scheme used in Fig. \ref{short}(a). The caustic intensity is the peak intensity in
the cutoff harmonics and the
cutoff energy is extracted for each time delay by sight. We note that there are
two cutoff energies as using a two-color field results in HHG spectra with
double plateaus \cite{double_plateau_static}. These two plateaus correspond to trajectories launched at a combined peak/peak
or peak/trough 
of the two color field. As a result, the caustic intensity and cutoff
energies
oscillate with half the period of the 400-nm pulse: The half-(800nm)-cycle
symmetry of the three-step HHG mechanism is broken by the inclusion of the
400~nm pulse. 

The two cutoff energies are anti-correlated\textemdash{} an
extension of the higher-energy cutoff reduces the cut-off energy of the first
plateau. At certain time delays the cutoff energies overlap,
and we observe a single plateau: for these time delays, classical trajectory
calculations reveal that there are 
trajectories of approximately equal return energy launched on multiple field
peaks. The peak caustic intensity occurs when the cutoff of the first plateau
is at a minimum.
The 
intensity along the arms of the swallowtail in Fig. \ref{short} then decreases
as the pulse arrangement yields higher-energy trajectories at the expense of a
lower ionization probability.

Extending the second plateau to higher energy comes at the expense of a
reduction of yield in the second plateau. As the cutoff energy increases, the caustic at the first
cutoff becomes more intense, dominating the spectrum. 
The peak caustic intensity occurs when
the second cutoff energy is at its highest. This occurs at a time delay of
0.17~fs,  not zero delay as might be expected naively. To elucidate this further, we perform calculations for
different primary wavelengths, and find that the offset from zero-delay is a constant
phase difference of approximately $\pi/7$ (Tab. \ref{peaktable}).

\begin{table}[b]
    \begin{tabular}{ c  c  c  c }
     \hline \hline \vspace{-5pt} \\
    Wavelength (nm) & Period (fs) & Offset (fs) & Phase offset (rad) \\ \hline \vspace{-7pt} \\
    600  & 2.00 & 0.14 & 0.44 \\
    800  & 2.67 & 0.17 & 0.39 \\
    1300 & 4.33 & 0.33 & 0.48 \\ \hline \hline
    \end{tabular}
   \caption{The delay time, which gives the peak harmonic emission
     and highest cut-off energy for neon in an $\omega- 2\omega$ pulse
     scheme, is an approximately constant phase offset for three different values of
     the primary ($\omega$) wavelength.
     \label{peaktable}
   }
\end{table}

\begin{figure}[!t]
  \centering
  \includegraphics[width=0.49\textwidth]{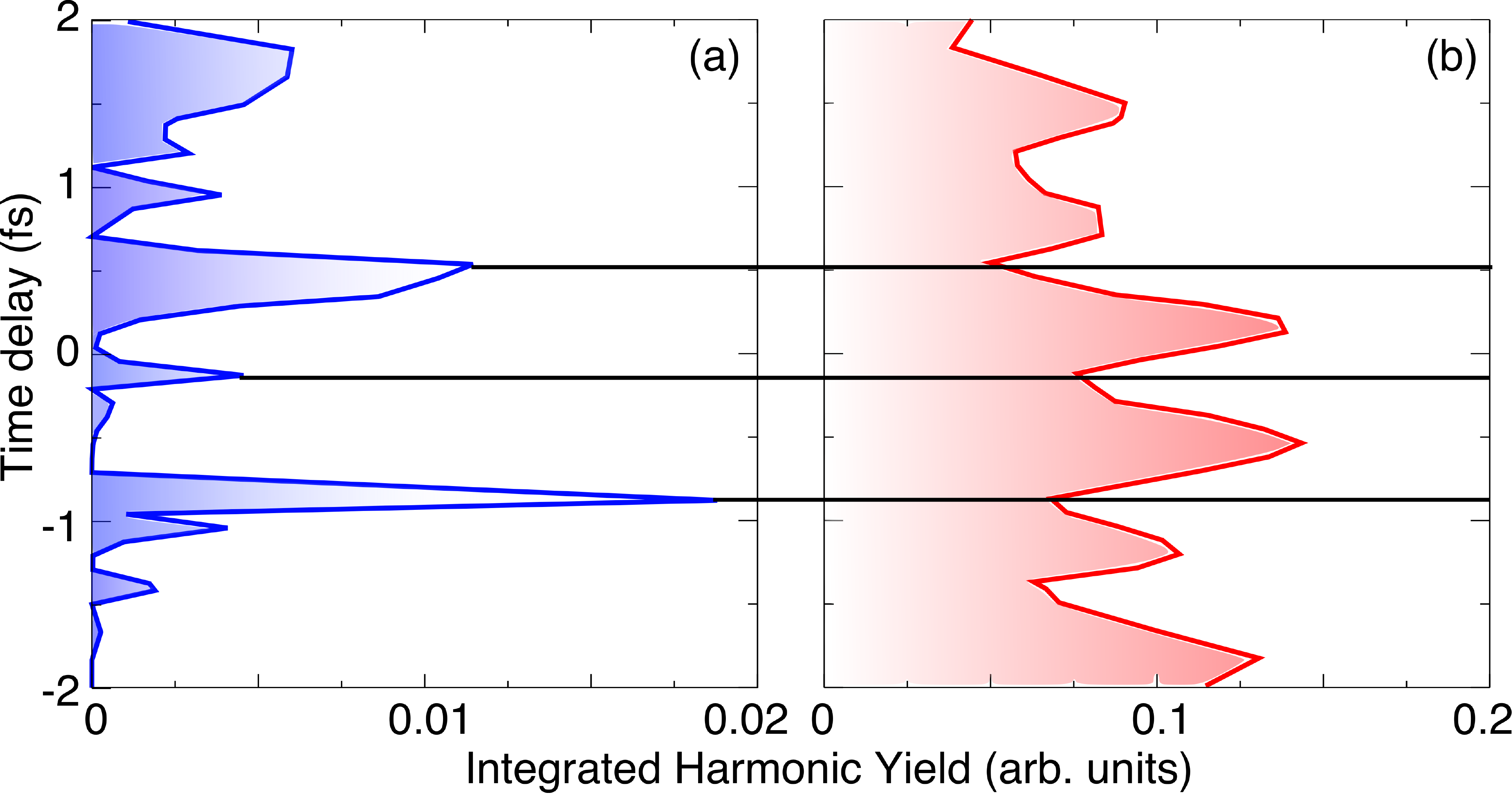}
  \caption{Integrated harmonic yield of the first (b) and second (a) plateaus
    arising in the harmonic spectra of neon exposed to two short laser pulses as
    described in Fig. \ref{short}a. The variation of the yield, as
    marked by the horizontal lines, is periodic with half the period of the
    400-nm pulse. \label{heights}}
\end{figure}

This offset, approximately 7\% of the field period, is the proportion of the
pulse peak where the electric field can be considered to be quasistatic (over
97\% of its peak strength). Tunnel ionization is most likely in this quasistatic
window.  For this optimal phase offset the peak of the $2\omega$ pulse arrives
just after that of the fundamental pulse, broadening the
central peak and enhancing ionization.  The green shaded regions of Fig.
\ref{trajectories} highlight the trajectories which are launched by such
broadened peak. It can be seen that these trajectories yield lower-energy
harmonics than those which are launched in the 
unshaded regions. Thus the most likely ionization event yields the spectral
caustic, while at the same time delay, much higher-energy trajectories are launched
but with vanishingly small probability (Fig. \ref{heights}). For the two
time delays on either side of zero delay which yield the most intense caustics, the
yield at the first plateau is three orders of magnitude larger than the yield in the
second. 

The main caustic, (time delay $\approx 0.17$fs ) has nearly twice the intensity
of the second (time delay $\approx 0.5$fs). This is because the field strength
at the central peak is higher than in the penultimate trough. Both of these
caustics arise when a peak (trough) of the $2\omega$ pulse arives just after a
peak (trough) of the $\omega$ pulse.
For a time delay one half cycle of the $2\omega$ pulse sooner or later, the
positions of the peaks and troughs of the 400-nm pulse relative to the 800-nm
pulse are reversed. A trough of the $2\omega$ pulse occuring after the peak of
the 800-nm pulse results in a shorter, flatter fundamental peak decreasing the
ionization probability and therefore the harmonic yield. Thus periodic reduction in
cumulative yield, seen as the blue horizontal bands in Fig. \ref{short}(a) and
more explicitly in Fig. \ref{heights}, occur
when a
trough of the 400-nm pulse coincides with the peak of the 800-nm pulse. 

\section{\label{sec:conclusion}Conclusion}

The continued improvement of both large-scale and table-top laser technology has
created increased flexibility to shape light pulses, controlling both their
spectral and temporal properties \cite{synthesised_light_transients}. Although
an aligned $\omega-2\omega$ scheme is by no means the most sophisticated or complex
arrangement, it nonetheless provides a tractable insight into the strong-field
dynamics of atomic electrons. In this paper we have assessed the behavior of a
neon atom exposed to such a combination of pulses. While it was previously well known
that employing the second harmonic field could substantially increase both the
harmonic yield and cut-off, we have seen that using a short, few-cycle $2\omega$
pulse can elicit more interesting behaviours.  In particular, a short $2\omega$ pulse
can stimulate different interference structures in the harmonic spectrum\textemdash{}
leading to spectral caustics and continua or comblike fringes depending on the
precise shape of the pulse.

Also, using both the RMT method and a classical trajectory model we are able to
assess the mechanisms by which the HHG yield and cut-off are affected. Namely,
a broadening of the fundamental peak leads to increased tunnel ionization
and thereby increased HHG yield, while a simultaneous narrowing of the previous trough increases the electron excursion time, increasing the recollision velocity and
hence the cut-off energy. 

With the enhanced ability to shape light pulses in experiment, it will become
increasingly important to understand in detail the response of atoms and
molecules to more
complex arrangements of laser pulses. To this end, it will be interesting to use
the capabilities of the RMT method to identify schemes whereby the signatures
of multielectron correlation and atomic structure can be extracted from high-order-harmonic spectra. 

\section{\label{sec:acknowlegements}Acknowledgements}
KH is supported by the Department for Employment and Learning Northern Ireland under the
programme for government.
HWH acknowledges
financial support from the Engineering and Physical Sciences Research Council under Grant No. EP/G055416/1 and the European Union Initial
Training Network CORINF. 
This work used the
ARCHER UK National Supercomputing Service (\url{archer.ac.uk}).
The data used in this paper may be found using
Ref. \cite{kathryn_neon_data}.

\bibliography{/users/abrown41/Documents/pub/bib_utils/mybib}

\end{document}